\documentclass[journal]{IEEEtran}
\ifCLASSINFOpdf

\usepackage{setspace}

\usepackage{cite}
\usepackage{soul}
\usepackage{caption}
\usepackage{subcaption}
\usepackage{array}
\usepackage{amsmath,amssymb,amsfonts}
\usepackage{setspace}
\usepackage{algorithmic}
\usepackage{graphicx}
\usepackage{tabularx}
\usepackage{booktabs}
\usepackage{setspace}
\usepackage{multirow}
\usepackage{fancyhdr}
\usepackage{float}
\usepackage{textcomp}
\usepackage{xcolor}
\usepackage{amsthm}
\usepackage{graphics}
\usepackage{algorithm}
\usepackage{algorithmic}
\usepackage{etoolbox}
\usepackage{cleveref}
\usepackage{subcaption}
\usepackage{booktabs}
\let\oldEquation\equation
\let\endOldEquation\endequation
\renewenvironment{equation}
  {\oldEquation\small}
  {\endOldEquation}
\begin{document}

\title{Movable Superdirective Pairs: A Phase Shifter-Free Approach to mmWave Communications}

\author{Liangcheng~Han,
        Haifan~Yin,~\IEEEmembership{Senior Member,~IEEE},~Mengying~Gao, and~Rui~Zhang,~\IEEEmembership{Fellow,~IEEE}
\thanks{Liangcheng Han, Haifan Yin, and Mengying Gao are with the School of Electronic Information and Communications, Huazhong University of Science and Technology, Wuhan 430074, China (e-mail: \{hanlc, yin, mengyinggao\}@hust.edu.cn). The corresponding author is Haifan Yin.}
\thanks{R. Zhang is with School of Science and Engineering, Shenzhen Research Institute of Big Data, The Chinese University of Hong Kong, Shenzhen, Guangdong 518172, China (e-mail: rzhang@cuhk.edu.cn). He is also with the Department of Electrical and Computer Engineering, National University of Singapore, Singapore 117583 (e-mail: elezhang@nus.edu.sg).}
\thanks{This work was supported by the Fundamental Research Funds for the Central Universities and the National Natural Science Foundation of China under Grant 62071191.}}

\maketitle

\begin{abstract}
In this letter, we propose a novel Movable Superdirective Pairs (MSP) approach that combines movable antennas with superdirective pair arrays to enhance the performance of millimeter-wave (mmWave) communications on the user side. By controlling the rotation angles and positions of superdirective antenna pairs, the proposed MSP approach maximizes the received signal-to-noise ratio (SNR) of multipath signals without relying on phase shifters or attenuators. This approach addresses the limitations of traditional superdirective antennas, which are typically restricted to the endfire direction and suffer from reduced scanning bandwidth and increased complexity. An efficient algorithm based on alternating optimization and the gradient projection method is developed to solve the non-convex optimization problem of antennas' joint rotating positioning. Simulation results demonstrate that the MSP approach achieves significant performance gains over fixed-position array (FPA) employing Maximum Ratio Combining (MRC), while reducing system complexity and hardware costs.
\end{abstract}

\begin{IEEEkeywords}
Millimeter-wave communications, superdirective antennas, movable antennas.
\end{IEEEkeywords}
\IEEEpeerreviewmaketitle
\vspace{-0.4cm}
\section{Introduction}
Millimeter-wave (mmWave) communications offer significantly higher data rates due to the abundant spectrum availability. However, mmWave channels suffer from severe path loss~\cite{niu2015survey}, necessitating the use of highly directive antennas to compensate for signal attenuation.

Superdirective antennas emerge as a promising solution, capable of achieving an $M^2$ directivity gain in the endfire direction when the antenna spacing approaches zero~\cite{uzkov1946approach}. Despite this advantage, superdirective antennas have several limitations. Increasing the number of elements in such arrays leads to heightened ohmic losses and greater sensitivity to excitation coefficient inaccuracies~\cite{han2024superdirective,gao2024robust}. Additionally, achieving proper impedance matching becomes more complex~\cite{nie2014systematic}, and the active port impedance varies with the excitation coefficients~\cite{hannan1965impedance}, reducing the scanning bandwidth.

To mitigate these issues, recent literature has proposed the use of superdirective pair structures~\cite{dovelos2023superdirective}. Each superdirective pair consists of two closely coupled elements, simplifying the matching scheme and reducing ohmic losses and sensitivity issues. Simulations have demonstrated the effectiveness of this configuration in enhancing communication energy efficiency. Further advancements were made in~\cite{lynch2024super}, where a two-element superdirective antenna array achieved a gain of 4.268 (considering matching and ohmic losses) at a spacing of $0.2\lambda$. However, these superdirective designs are limited to providing high gain only in the endfire direction.

Moreover, mmWave radio frequency (RF) devices are expensive and introduce higher losses compared to their sub-6\,GHz counterparts~\cite{song2021terahertz}. Traditional beamforming techniques that employ phase shifters and attenuators can further exacerbate these costs and losses.

Recently, movable antenna (MA) and six-dimensional movable antenna (6DMA) have attracted significant interest as they introduce new degrees of freedom in antenna manipulation, aiding in the improvement of wireless communication performance~\cite{zhu2023modeling}\cite{shao2024movable}. In this letter, we propose a novel approach that combines movable antennas with superdirective pair arrays, termed Movable Superdirective Pairs (MSP). By controlling the rotation angle and the $y$-axis position of the superdirective pairs, we aim to maximize the received signal-to-noise ratio (SNR) of multipath signals without the need for phase shifters. The advantages of the proposed approach include the preservation of scanning bandwidth and the ability to achieve high gain beyond the endfire direction, thus addressing the limitations of existing superdirective designs.

The optimization of the rotation angles and positions constitutes a highly non-convex problem. To tackle this challenge, we develop an efficient algorithm based on alternating optimization combined with gradient projection method. Simulation results demonstrate the proposed movable superdirective antenna pair approach achieves higher spectral efficiency than fixed-position antenna (FPA) arrays even with Maximum Ratio Combining (MRC).

\textit{Notations}: Throughout this letter, scalars are denoted by italic letters, e.g., $a$, vectors by boldface lowercase letters, e.g., $\mathbf{a}$, and matrices by boldface uppercase letters, e.g., $\mathbf{A}$. The transpose and Hermitian transpose operators are represented by $(\cdot)^T$ and $(\cdot)^H$, respectively. The notation $\mathbb{E}[\cdot]$ denotes the expectation operator. The real and imaginary parts of a complex number are denoted by $\text{Re}\{\cdot\}$ and $\text{Im}\{\cdot\}$, respectively. The Euclidean norm of a vector is denoted by $\|\cdot\|$. 

\section{System Model and Problem Formulation}
\subsection{System Model}
\begin{figure}[t]
  \centering
  \includegraphics[width=2.8in]{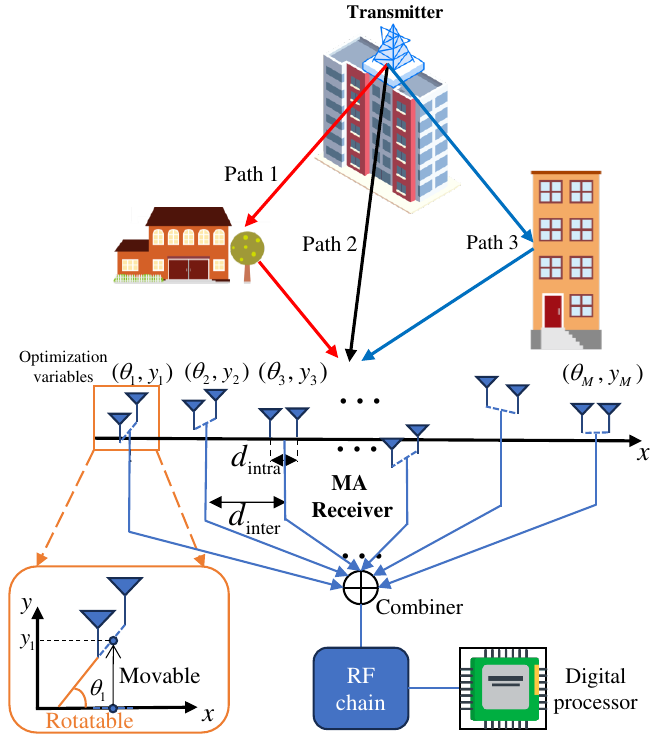}  
  \caption{Illustration of the MSP communication system.}\label{model}
  \vspace{-0.5cm}
\end{figure}
As illustrated in Fig.~\ref{model}, we consider a wireless communication system where a base station (BS) equipped with $N$ antennas communicates with a receiver in a multipath environment.  The movable receive array consists of $2M$ elements, grouped into $M$ antenna pairs. Each pair can rotate by an angle $\theta_i$ around its center point and move along the $y$-axis to position $y_i$. The intra-pair spacing $d_{\text{intra}}$ between the two antennas within each pair is very small (approximately $0.2\lambda$), resulting in strong coupling and thus achieving superdirectivity. The inter-pair spacing $d_{\text{inter}}$ between adjacent pairs is relatively large (typically not less than $0.5\lambda$) to reduce coupling between antenna pairs, and the pairs are arranged along the $x$-axis.

The center position of the $i$-th antenna pair is given by
\begin{equation}
\mathbf{r}_{ic} = \left[ (i-1)d_{\text{inter}},\, y_i \right]^T.
\end{equation}
We define the relative position vector between the two antennas within each pair as
\begin{equation}
\footnotesize
\Delta \mathbf{r} = \dfrac{d_{\text{intra}}}{2} \begin{pmatrix} \cos \theta_i \\ \sin \theta_i \end{pmatrix}.
\end{equation}
The positions of the two antennas within the $i$-th pair are
\begin{equation}
\mathbf{r}_{in} = \mathbf{r}_{ic} + (-1)^n\Delta\mathbf{r},\,n=1,2.
\end{equation}
The steering vector for the $i$-th antenna pair in the direction $\phi$ is defined as
\begin{equation}
\mathbf{a}_i(\phi) = [
e^{j k \mathbf{r}_{i1}^T \mathbf{u}(\phi)},\,
e^{j k \mathbf{r}_{i2}^T \mathbf{u}(\phi)}]^T,
\end{equation}
where $k = \dfrac{2\pi}{\lambda}$ is the wave number, and $\mathbf{u}(\phi) = [\cos \phi,\, \sin \phi]^T$ is the unit vector in the direction $\phi$.

Considering mutual coupling between the two antennas within each pair, the mutual impedance matrix of the $i$-th antenna pair is given by
\begin{equation}
\mathbf{Z}_0 = \begin{pmatrix}
Z_{\text{self}} & Z_{\text{mutual}} \\
Z_{\text{mutual}} & Z_{\text{self}}
\end{pmatrix},
\end{equation}
where $Z_{\text{self}} = R_{\text{self}} + j X_{\text{self}}$ is the self-impedance, and $Z_{\text{mutual}} = R_{\text{mutual}} + j X_{\text{mutual}}$ is the mutual impedance between the two antennas. Due to the sufficiently large inter-pair spacing, mutual coupling between different antenna pairs is negligible, rendering the total mutual impedance matrix $\mathbf{Z}\in\mathbb{C}^{2M\times 2M}$ of the array block-diagonal, i.e., 
\begin{equation}
\mathbf{Z} = \text{diag}\left( \mathbf{Z}_0,\, \mathbf{Z}_0,\, \dots,\, \mathbf{Z}_0 \right).
\end{equation}

To achieve maximum radiation in the end-fire direction $\theta_i$, the optimal excitation currents for each antenna pair under mutual coupling are derived as \cite{dovelos2023superdirective}
\begin{equation} \label{eq:i_opt}
\mathbf{i}_i^{\text{opt}} = \sqrt{ \dfrac{2P_t}{ \tilde{\mathbf{a}}^H(\theta_i) \left( \text{Re}\{\mathbf{Z}_0\} \right)^{-1} \tilde{\mathbf{a}}(\theta_i) } } \left( \text{Re}\{\mathbf{Z}_0\} \right)^{-1} \tilde{\mathbf{a}}^*(\theta_i),
\end{equation}
where $P_t$ is the total input power allocated to the antenna pair. To ensure that the optimal excitation coefficients do not change with the antenna pair position, we introduce the relative steering vector for each antenna pair, defined as
\begin{equation}
\tilde{\mathbf{a}}(\phi; \theta_i) = 
[e^{j k \Delta \mathbf{r}^T \mathbf{u}(\phi)} ,\,
e^{- j k \Delta \mathbf{r}^T \mathbf{u}(\phi)}]^T.
\end{equation}
In \cite{dovelos2023superdirective}, it has been proven that the amplitude of each excitation coefficient needs to be equal. 
This characteristic allows the antenna pair to be connected by a single cable with equal amplitude excitation, simplifying the hardware design. The phase difference between  currents can be managed using parasitic components, while maintaining equal amplitudes ensures efficient power utilization.
The radiation pattern for the $i$-th antenna pair is expressed in terms of the optimal currents as
\begin{equation}
F_i(\phi;\theta_i,y_i) = \mathbf{a}_i^T(\phi)\, \mathbf{i}_i^{\text{opt}},
\end{equation}
 Substituting \eqref{eq:i_opt} into the above expression yields
     \begin{equation}
     \footnotesize
F_i(\phi;\theta_i,y_i) = \sqrt{ \frac{2P_t}{ \mathbf{a}_i^H(\theta_i) \left( \text{Re}\{\mathbf{Z}_0\} \right)^{-1} \mathbf{a}_i(\theta_i) } }\mathbf{a}_i^T(\phi)\, \left( \text{Re}\{\mathbf{Z}_0\} \right)^{-1}\tilde{\mathbf{a}}^*(\theta_i),
 \end{equation}
indicating that the radiation pattern is influenced by mutual coupling through $\left( \text{Re}\{\mathbf{Z}_0\} \right)^{-1}$.

\vspace{-0.35cm}


\subsection{Problem Formulation}
We consider a multipath channel with $ L $ paths, where  path $ l $  has an angle of departure (AoD) $ \theta_l^{\text{BS}} $ from the BS, and an angle of arrival (AoA) $ \phi_l $ at the receiver. The BS is equipped with an $ N $-antenna array and applies beamforming using a fixed precoding scheme such as Zero Forcing (ZF). This beamforming enhances the signal strength in specific directions, resulting in an angle-dependent gain $ G_{\text{BS}}(l) $.   We introduce a path-dependent loss coefficient $ \beta_l $, representing propagation loss, shadowing, and small-scale fading effects. The effective signal strength for path $ l $ is defined as:
\begin{equation}
A_l = G_{\text{BS}}(l) \beta_l,
\end{equation}

Let $x$ be the transmitted symbol with zero mean and $\mathbb{E}(x^2)=1$. The received signal at the $ i $-th antenna pair, including additive white Gaussian noise (AWGN) $ n_i \sim \mathcal{CN}(0, \sigma^2) $, is given by
\begin{equation}
s_i = \sum_{l=1}^{L} A_l  F_i(\phi_l; \theta_i,y_i)  x + n_i.
\end{equation}
The total received signal combining all antenna pairs is
\begin{equation}
y = \sum_{i=1}^{M} s_i = \left( \sum_{l=1}^{L} A_l \sum_{i=1}^{M} F_i(\phi_l; \theta_i,y_i) \right) x + \sum_{i=1}^{M} n_i.
\end{equation}
Assuming the noise terms $ n_i $'s are independent and identically distributed, the aggregate noise has power $ \sigma_{\text{total}}^2 = M \sigma^2 $.

The signal power is
\begin{equation}
P_{\text{signal}} = \left| \sum_{i=1}^{M} \sum_{l=1}^{L} A_l\, F_i(\phi_l;\theta_i,y_i) \right|^2.
\end{equation}

We aim to maximize the SNR at the  combiner output and the  optimization problem is formulated as follows:
\begin{equation}
    \begin{aligned}
\text{(P1)} \quad \max_{\{ \theta_i,\, y_i \}} \quad & \text{SNR} = \dfrac{ \left| \displaystyle \sum_{i=1}^{M} \sum_{l=1}^{L} A_l\, F_i(\phi_l;\theta_i,y_i) \right|^2 }{ M \sigma^2 }, \\
\text{subject to} \quad & \theta_i \in [0,\, 2\pi), \quad y_i \in [y_{\min},\, y_{\max}], \quad \forall\, i.
\end{aligned}
\end{equation}

\textbf{Remark:} By controlling the rotation angles $ \theta_i $ and the $ y $-axis positions $ y_i $ of the antenna pairs, we can exploit the superdirectivity effect achieved by the antenna pairs in the endfire direction to enhance the overall array gain. This approach allows us to eliminate the dependence on phase shifters. However, the radiation pattern expression $ F_i(\phi_l;\theta_i,y_i) $ is quite complex, with the optimization variables $ \theta_i $ and $ y_i $ located inside exponential terms and coupled with each other. This coupling leads to a highly non-convex optimization problem.

\vspace{-0.25cm}
\section{Proposed Algorithms for (P1)}

In this section, we propose algorithms to solve the optimization problem (P1). Due to the coupling between the rotation angles $ \theta_i $ and the $ y $-axis positions $ y_i $, directly optimizing them jointly is complex. We employ an Alternating Optimization (AO) algorithm combined with the Gradient Projection Method (GPM) to iteratively optimize $ \theta_i $ and $ y_i $.
\vspace{-0.3cm}
\subsection{Alternating Optimization Framework}

The AO algorithm alternates between optimizing $ \theta_i $ and $ y_i $:
\begin{enumerate}
    \item  Keeping $ y_i $ constant, we optimize $ \theta_i $.
    \item  Keeping $ \theta_i $ constant, we optimize $ y_i $.
\end{enumerate}
In each step, we apply the GPM, which requires computing the gradients of the objective function regarding $ \theta_i $ and $ y_i $.
\vspace{-0.3cm}
\subsection{Gradient Computation}
The key to  GPM is computing gradients $ \frac{\partial P_{\text{signal}}}{\partial \theta_i} $ and $ \frac{\partial P_{\text{signal}}}{\partial y_i} $.

\subsubsection{Gradient with Respect to $ \theta_i $}

We first compute the gradient of $ P_{\text{signal}} $ with respect to $ \theta_i $.

The total received power is:
\begin{equation}
P_{\text{signal}} = |S_{\text{total}}|^2 = S_{\text{total}} S_{\text{total}}^*,
\end{equation}
where
\begin{equation}
S_{\text{total}} = \sum_{i=1}^{M} s_i = \sum_{i=1}^{M} \sum_{l=1}^{L} A_l\, F_i(\phi_l;\theta_i,y_i).
\end{equation}

The gradient of $ P_{\text{signal}} $ with respect to $ \theta_i $ is
\begin{equation}
     \begin{aligned}
\frac{\partial P_{\text{signal}}}{\partial \theta_i} = \frac{\partial}{\partial \theta_i} \left( S_{\text{total}} S_{\text{total}}^* \right)  
= 2\, \text{Re} \left\{ \left( \frac{\partial S_{\text{total}}}{\partial \theta_i} \right) S_{\text{total}}^* \right\}.
\end{aligned}
\end{equation}

Since $ S_{\text{total}} $ depends on $ \theta_i $ through $ F_i(\phi_l;\theta_i,y_i) $, we have:
\begin{equation}
\frac{\partial S_{\text{total}}}{\partial \theta_i} = \frac{\partial s_i}{\partial \theta_i} = \sum_{l=1}^{L} A_l\, \frac{\partial F_i(\phi_l;\theta_i,y_i)}{\partial \theta_i}.
\end{equation}

Differentiating $ F_i(\phi_l;\theta_i,y_i) $ with respect to $ \theta_i $ yields
\begin{equation}
\begin{aligned}
\frac{\partial F_i(\phi_l; \theta_i, y_i)}{\partial \theta_i} = 
\frac{\partial \mathbf{a}_i^T(\phi_l)}{\partial \theta_i} 
\left( \text{Re}\{\mathbf{Z}_0\} \right)^{-1}\tilde{\mathbf{a}}^*(\theta_i).
\end{aligned}
\end{equation}

Each element of $ \mathbf{a}_i(\phi_l) $ depends on $ \theta_i $ through
\begin{equation}
\left[ \mathbf{a}_i(\phi_l) \right]_n = e^{j k \mathbf{r}_{in}^T \cdot \mathbf{u}(\phi_l)}.
\end{equation}
Differentiating $\mathbf{a}_i(\phi_l)$ with respect to $ \theta_i $ yields
\begin{equation}
    \frac{\partial \mathbf{a}_i(\phi_l)}{\partial \theta_i} = j k \begin{bmatrix}
\frac{\partial (\mathbf{r}_{i1}^T \mathbf{u}(\phi_l))}{\partial \theta_i} e^{j k \mathbf{r}_{i1}^T \mathbf{u}(\phi_l)} \\
\frac{\partial (\mathbf{r}_{i2}^T \mathbf{u}(\phi_l))}{\partial \theta_i} e^{j k \mathbf{r}_{i2}^T \mathbf{u}(\phi_l)}
\end{bmatrix}.
\end{equation}
Differentiating $ \mathbf{r}_{in},n=1,2 $ with respect to $ \theta_i $ yields
\begin{equation}
\frac{\partial \mathbf{r}_{in}}{\partial \theta_i} = \frac{d_{\text{intra}}}{2} (-1)^{n} \begin{bmatrix}
-\sin \theta_i \\
\cos \theta_i
\end{bmatrix}.
\end{equation}
Thus,
\begin{equation}
    \begin{aligned}
\frac{\partial (\mathbf{r}_{in}^T \mathbf{u}(\phi_l))}{\partial \theta_i} 
&= \frac{\partial \mathbf{r}_{in}}{\partial \theta_i}^T \mathbf{u}(\phi_l) \\
&= \frac{d_{\text{intra}}}{2} (-1)^{n} 
    \begin{bmatrix}
        -\sin \theta_i & \cos \theta_i
    \end{bmatrix}
    \begin{bmatrix}
        \cos \phi_l \\
        \sin \phi_l
    \end{bmatrix} \\
&= \frac{d_{\text{intra}}}{2} (-1)^{n} 
     \sin(\phi_l - \theta_i).
\end{aligned}
\end{equation}
Substituting back, we obtain:
\begin{equation}
    \begin{aligned}
    \frac{\partial \mathbf{a}_i(\phi_l)}{\partial \theta_i} 
    = j k \frac{d_{\text{intra}}}{2} \sin(\phi_l - \theta_i)
    \begin{bmatrix}
         -e^{j k \mathbf{r}_{i1}^T \mathbf{u}(\phi_l)} \\
         e^{j k \mathbf{r}_{i2}^T \mathbf{u}(\phi_l)}
    \end{bmatrix}.
\end{aligned}
\end{equation}
\subsubsection{Gradient with Respect to $ y_i $}
Next, we compute $ \frac{\partial P_{\text{signal}}}{\partial y_i} $.
\begin{equation}
\frac{\partial S_{\text{total}}}{\partial y_i} = \frac{\partial s_i}{\partial y_i} = \sum_{l=1}^{L} A_l\, \frac{\partial F_i(\phi_l;\theta_i,y_i)}{\partial y_i}.
\end{equation}
Since $ y_i $ affects the $ y $-coordinate of $ \mathbf{r}_{in} $, we have:
    \begin{equation}
\frac{\partial F_i(\phi_l;\theta_i,y_i)}{\partial y_i} = \left( \frac{\partial \mathbf{a}_i^T(\phi_l)}{\partial y_i} \right)  \left( \text{Re}\{\mathbf{Z}_0\} \right)^{-1} \tilde{\mathbf{a}}^*(\theta_i).
\end{equation}
Differentiating with respect to $ y_i $:
\begin{equation}
\frac{\partial \left[ \mathbf{a}_i(\phi_l) \right]_n}{\partial y_i} = j k \sin \phi_l\, e^{j k \mathbf{r}_{in}^T \cdot \mathbf{u}(\phi_l)}.
\end{equation}
Therefore,
\begin{equation}
\frac{\partial F_i(\phi_l;\theta_i,y_i)}{\partial y_i} = j k \sin \phi_l\, F_i(\phi_l;\theta_i).
\end{equation}
The gradient can thus be calculated as:
\begin{equation}
        \begin{aligned}
\frac{\partial P_{\text{signal}}}{\partial y_i} = -2 k \sum_{l=1}^{L} A_l \sin \phi_l\, \text{Im} \left\{ F_i(\phi_l;\theta_i,y_i)\, S_{\text{total}}^* \right\}.
\end{aligned}   
\end{equation}
\vspace{-0.8cm}
\subsection{Adam Optimizer for Learning Rate Control}

To efficiently update $ \theta_i $ and $ y_i $, we utilize the Adam optimizer \cite{kingma2014adam}, which adapts the learning rates for each parameter based on the first and second moments of the gradients. This method accelerates convergence and improves stability.
For each parameter $ \theta_i $, the Adam optimizer updates are as follows:

    \begin{equation}
    \footnotesize
          \begin{aligned}
m_{\theta_i}^{(k)} &= \beta_1 m_{\theta_i}^{(k-1)} + (1 - \beta_1) \frac{\partial P_{\text{signal}}}{\partial \theta_i}, \\
v_{\theta_i}^{(k)} &= \beta_2 v_{\theta_i}^{(k-1)} + (1 - \beta_2) \left( \frac{\partial P_{\text{signal}}}{\partial \theta_i} \right)^2, \\
\hat{m}_{\theta_i}^{(k)} &= \frac{m_{\theta_i}^{(k)}}{1 - \beta_1^k}, 
\hat{v}_{\theta_i}^{(k)} = \frac{v_{\theta_i}^{(k)}}{1 - \beta_2^k}, \\
\theta_i^{(k+1)} &= \theta_i^{(k)} - \alpha \frac{\hat{m}_{\theta_i}^{(k)}}{\sqrt{\hat{v}_{\theta_i}^{(k)}} + \epsilon},
\end{aligned} 
\end{equation}
and the update of $ y_i $ is similar.
Here, $ m_{\theta_i} $ and $ v_{\theta_i} $ are the first and second moment estimates for $ \theta_i $, $ \beta_1 $ and $ \beta_2 $ are exponential decay rates for the moment estimates, typically $ \beta_1 = 0.9 $ and $ \beta_2 = 0.999 $, $ \epsilon $ is a small constant to prevent division by zero (e.g., $ \epsilon = 10^{-8} $), and $ \alpha $ is the learning rate.
After each update, variables are projected onto their feasible sets according to
\begin{equation}
\footnotesize
       \begin{aligned}
\theta_i^{(k+1)} &\leftarrow \theta_i^{(k+1)} \mod 2\pi, \\
y_i^{(k+1)} &\leftarrow \min\left( \max\left( y_i^{(k+1)}, y_{\min} \right), y_{\max} \right).
\end{aligned} 
\end{equation}

\begin{algorithm}[t]
\small
\caption{Alternating Optimization for Problem (P1)}
\label{alg:proposed}
\begin{algorithmic}[1]
\STATE \textbf{Initialization:} Set initial \( \theta_i^{(0)} \) and \( y_i^{(0)} \) within their feasible ranges. Initialize the first and second moment estimates \( m_{\theta_i}^{(0)} = 0 \), \( v_{\theta_i}^{(0)} = 0 \), \( m_{y_i}^{(0)} = 0 \), and \( v_{y_i}^{(0)} = 0 \).
\REPEAT
    \STATE \textbf{Optimize \( \theta_i \) with fixed \( y_i \):}
    \FOR{each \( i = 1, \ldots, M \)}
        \STATE Compute \( \frac{\partial P_{\text{signal}}}{\partial \theta_i} \).
        \STATE Update \( \theta_i \) using the Adam optimizer.
        \STATE Project \( \theta_i \) onto the feasible set.
    \ENDFOR
    \STATE \textbf{Optimize \( y_i \) with fixed \( \theta_i \):}
    \FOR{each \( i = 1, \ldots, M \)}
        \STATE Compute \( \frac{\partial P_{\text{signal}}}{\partial y_i} \).
        \STATE Update \( y_i \) using the Adam optimizer.
        \STATE Project \( y_i \) onto the feasible set.
    \ENDFOR
\UNTIL{convergence}
\STATE \textbf{Output:} Optimized \( \theta_i \) and \( y_i \) for all \( i \).
\end{algorithmic}
\end{algorithm}

The proposed algorithm is summarized in Algorithm~\ref{alg:proposed}. It decouples the joint optimization of $\theta_i$ and $y_i$ by iteratively fixing one set of variables while optimizing the other, which facilitates convergence to a stationary point of the objective function. However, due to the non-convex nature of the problem, convergence to a global optimum cannot be guaranteed. The  complexity per iteration is primarily dictated by the gradient computations of $P_{\text{signal}}$ with respect to $\theta_i$ and $y_i$, and the complexity order is $\mathcal{O}(ML)$.
\vspace{-0.45cm}
\section{Numerical results}
\vspace{-0.1cm}
\begin{spacing}{0.9}
    In this section, we present numerical simulations to evaluate the performance of the proposed MSP approach. The simulation parameters are set as follows: the intra-pair spacing is $d_{\text{intra}} = 0.2\lambda$, and the inter-pair spacing is $d_{\text{inter}} = 0.5\lambda$. The movable range along the $y$-axis for each antenna pair is constrained between $y_{\min} = -\lambda$ and $y_{\max} = \lambda$. The noise power is assumed to be $\sigma^2 = 1$. The mutual impedance between two isotropic antennas can be calculated as $\text{sinc}(\frac{kd}{\pi})$, where $d$ is the antenna spacing\cite{han2024superdirective}. Adjusting the positions along the $y$-axis offers limited capability for phase compensation of incoming waves along the $x$-axis. To mitigate this limitation, we set the initial phase difference between adjacent antenna pairs to $\pi$. This initial phase shift can be easily implemented using microstrip lines. Furthermore, we assume that the AoA of the incoming signals are confined to the x-y plane. To enhance the performance of the optimization algorithm, the initial rotation angles $\theta_i$, for $i = 1,\ldots,M$, are set close to the angle of the path with the largest amplitude. In the simulations, the amplitudes of the multipath components $ A_l $ are normalized such that  $ \sum_{l=1}^{L} |A_l|^2 = 1 $.
\end{spacing}

\begin{figure}[htbp]
    \centering
    \begin{subfigure}[b]{0.45\linewidth}
        \centering
        \includegraphics[width=\linewidth]{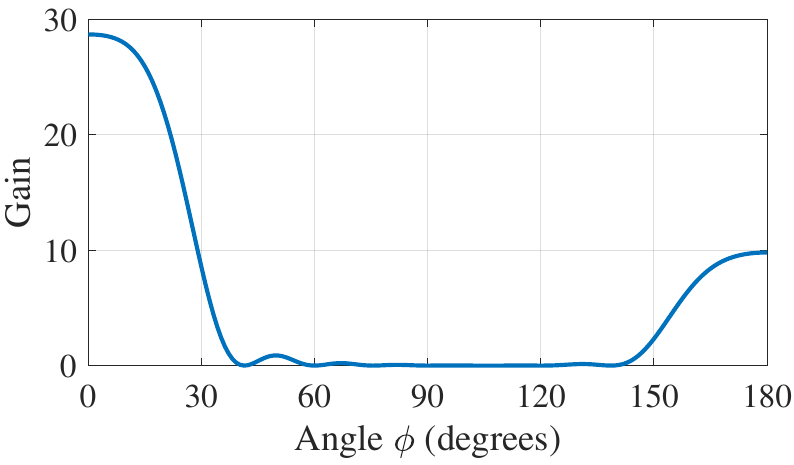}
        \caption{Initial pattern}
        \label{fig:init_pattern}
    \end{subfigure}
    \hfill
    \begin{subfigure}[b]{0.45\linewidth}
        \centering
        \includegraphics[width=\linewidth]{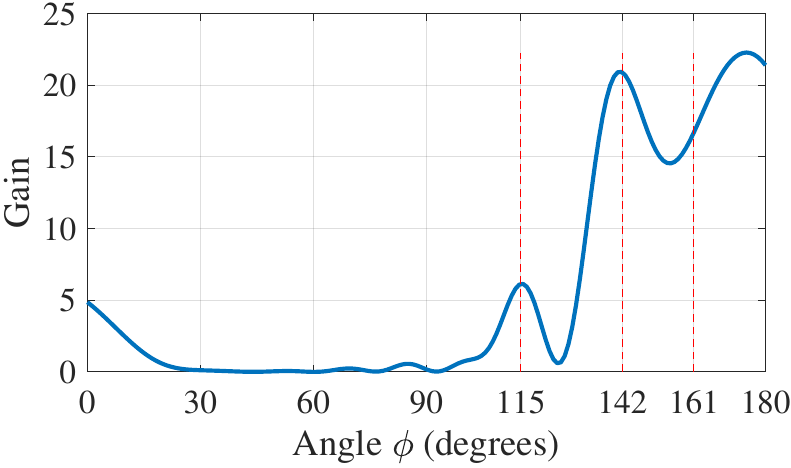}
        \caption{Optimized pattern}
        \label{fig:opt_pattern}
    \end{subfigure}
    \vspace{0cm}
    
    \begin{subfigure}[b]{0.45\linewidth}
        \centering
        \includegraphics[width=\linewidth]{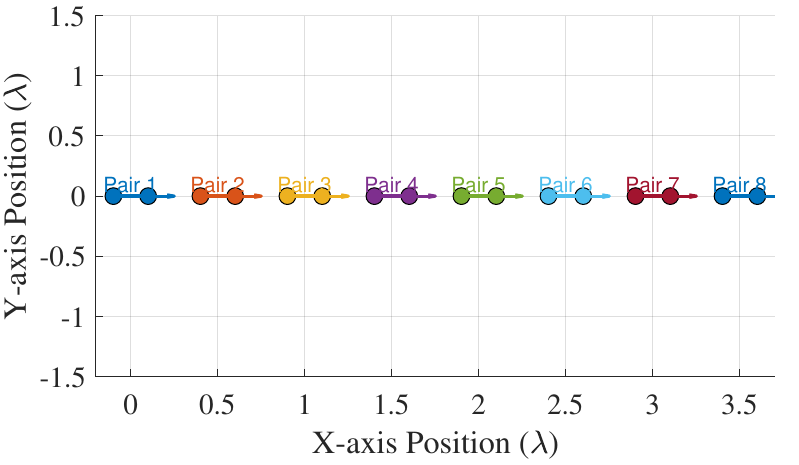}
        \caption{Initial array}
        \label{fig:init_pos}
    \end{subfigure}
    \hfill
    \begin{subfigure}[b]{0.45\linewidth}
        \centering
        \includegraphics[width=\linewidth]{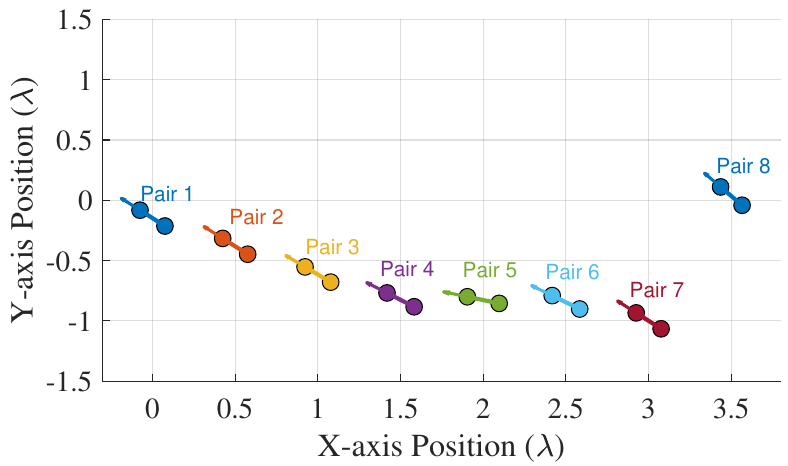}
        \caption{Optimized array}
        \label{fig:opt_pos}
    \end{subfigure}
    
    \caption{Comparison of initial and optimized patterns and array configurations.}
    \label{fig:comparison}
    \vspace{-0.5cm}
\end{figure}

Fig.~\ref{fig:comparison} illustrates the initial and optimized array patterns and configurations. Specifically, for $M = 8$ antenna pairs, we consider a multipath environment with three paths. The amplitudes and angles of the paths are given by $[0.4, 0.85, 0.32]$ and $[115^\circ, 142^\circ, 161^\circ]$. Fig.~\ref{fig:comparison}(\subref{fig:init_pattern}) and \ref{fig:comparison}(\subref{fig:opt_pattern}) show the radiation patterns before and after optimization, while Fig.~\ref{fig:comparison}(\subref{fig:init_pos}) and~\ref{fig:comparison}(\subref{fig:opt_pos}) depict the corresponding antenna array configurations.
From Fig.~\ref{fig:comparison}(\subref{fig:opt_pattern}), we observe that the optimized pattern exhibits significant radiation intensities in the directions of the multipath components. This demonstrates the effectiveness of the proposed algorithm in steering the superdirective gain towards the desired signal paths. 

\begin{figure}[htbp]
  \centering
  \includegraphics[width=2.5in]{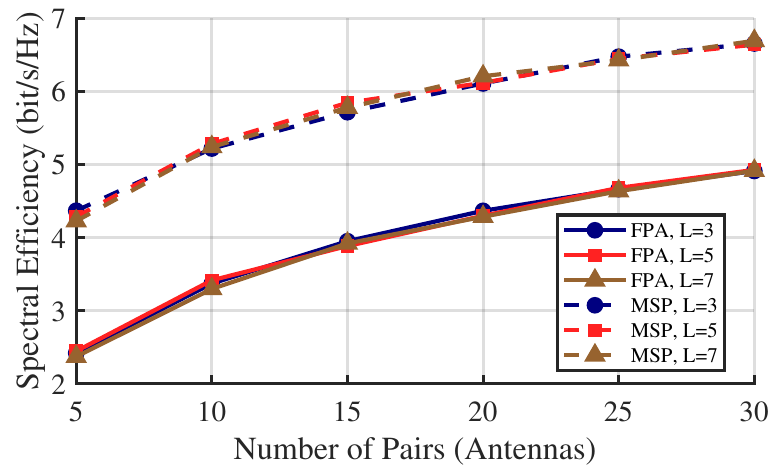}  
  \caption{Spectral efficiency comparison between the proposed MSP structure and the FPA with MRC reception under different numbers of antennas and paths.}
  \label{fig:compare2FPA}
  \vspace{-0.3cm}
\end{figure}

\begin{spacing}{0.9}
In addition, we conduct a simulation to compare the spectral efficiency of the proposed MSP structure with the FPA under different numbers of antennas. In this simulation, the FPA employs MRC beamforming at the receiver. For each number of antenna pairs $M$, we simulate three different numbers of multipath components: $L = 3$, $5$, and $7$. For each case, we perform 100 random realizations of the path angles and amplitudes, and calculate the average spectral efficiency.
Fig.~\ref{fig:compare2FPA} shows the simulation results, where the spectral efficiency is plotted against the number of antenna pairs $M$. It can be observed that the proposed MSP structure consistently outperforms the FPA by approximately $2$ bit-per-second-per-Hertz (bit/s/Hz)  in spectral efficiency across different numbers of antennas and paths. This improvement is attributed to the capability of MSP to adjust antenna positions and orientations to enhance the received signal strength in the directions of the multipath components. It is noteworthy that the spectral efficiency curves in Fig.~\ref{fig:compare2FPA} exhibit minimal sensitivity to the number of multipath components  $L$. This  insensitivity can be attributed to the sparse nature of mmWave channels, where a few dominant paths carry the majority of the signal energy.  The proposed MSP approach optimizes the antenna array configuration to maximize the gain towards these dominant paths.
Moreover, the proposed MSP approach avoids the insertion losses, typically around 5-6 dB, associated with phase shifters required in the FPA. By eliminating these components, the actual performance gain of the MSP would be even more significant when considering practical implementation losses. 

\begin{figure}[htbp]
  \centering
  \includegraphics[width=2.2in]{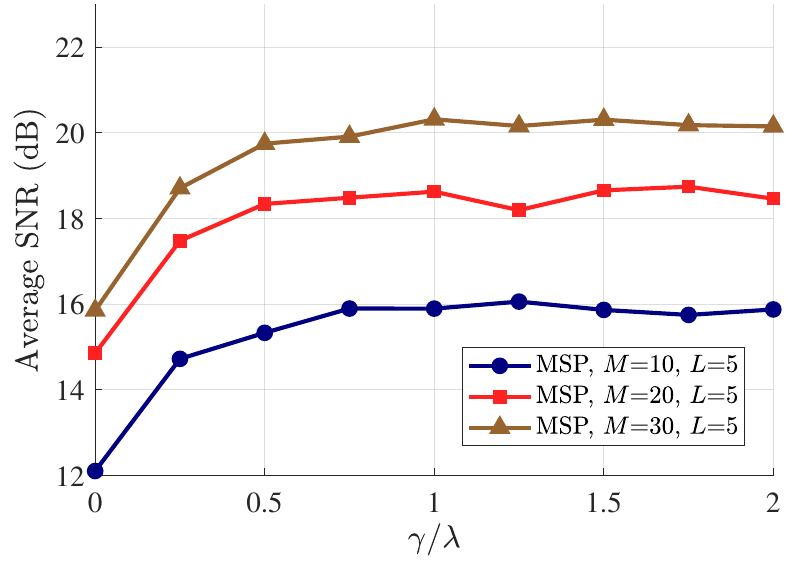}  
  \caption{Average SNR versus normalized movement range $ \gamma / \lambda $ for different numbers of antenna pairs $ M $.}
  \label{fig:SNR_vs_gamma}
  \vspace{-0.3cm}
\end{figure}

    We then investigate the effect of the antenna movement range along the $ y $-axis on the average received SNR. Specifically, we consider the movement range $ y_i \in [-\gamma,\, \gamma],i=1,\cdots,M $, where $ \gamma $ varies from $ 0 $ to $ 1.5\lambda $. We simulate the MSP system with $ L = 5 $ and  $ M = 10,\, 20,\, \text{and}\ 30 $.
Fig.~\ref{fig:SNR_vs_gamma} illustrates the average SNR versus the normalized movement range $ \gamma / \lambda $ for different numbers of antenna pairs. It can be observed that for each number of antennas, the average SNR increases as $ \gamma $ increases from zero. This indicates that a larger movement range allows the antenna pairs to better adjust their positions along the $ y $-axis to maximize the received signal power from the multipath components. However, when $ \gamma $ reaches $ \lambda $, the performance gain saturates, and the average SNR no longer increases with further increases in the movement range. The saturation of performance at $ \gamma = \lambda $ can be attributed to the periodicity of the array factor with respect to spatial displacement. Beyond a movement range of one wavelength, the antenna pairs may be  capable of covering the entire spatial angle range necessary to capture the multipath components effectively. 
\end{spacing}

To further evaluate the efficiency and effectiveness of the proposed algorithm, we conduct simulations to compare its runtime and performance with the Particle Swarm Optimization (PSO) algorithm. As shown in Table~\ref{tab:comparison}, the proposed algorithm achieves computational times approximately two orders of magnitude shorter than those attained by the PSO algorithm. Specifically, for $M=20$, the average runtime of the proposed algorithm is 0.4233 seconds, which is significantly less than the 22.9288 seconds required by PSO. When $M=30$, the proposed algorithm completes in an average of 0.5868 seconds, compared to PSO’s 45.1724 seconds. This substantial reduction in runtime clearly demonstrates the high efficiency of the proposed approach.
In terms of performance, the proposed algorithm not only maintains comparable average received SNR but even slightly outperforms the PSO algorithm. For $M=20$, the proposed algorithm attains an average SNR of 18.4641 dB, slightly higher than PSO's 17.9592 dB. When the number of antenna elements increases to $M=30$, the proposed algorithm achieves an average SNR of 20.2608 dB, surpassing PSO's 19.4728 dB.
These results indicate that as the number of antenna elements $M$ increases and the solution space becomes larger and more complex, the performance advantage of the proposed algorithm over PSO becomes more pronounced.
\begin{table}[ht]
\vspace{-0.2cm}
\centering
\caption{Comparison of algorithm performance for different configurations}
\label{tab:comparison}
\resizebox{\linewidth}{!}{%
\begin{tabular}{lcccc}
\toprule
Configuration & \multicolumn{2}{c}{$M$=20, $L$=5} & \multicolumn{2}{c}{$M$=30, $L$=5} \\
\cmidrule(r){2-3} \cmidrule(r){4-5}
Algorithm & Runtime (s) & SNR (dB) & Runtime (s) &  SNR (dB) \\
\midrule
Proposed Algorithm & \textbf{0.4233} & \textbf{18.4641} & \textbf{0.5868} & \textbf{20.2608} \\
PSO & 22.9228 & 17.9592 & 45.1724 & 19.4728
 \\
\bottomrule
\end{tabular}%
}
\vspace{-0.3cm}
\end{table}
\vspace{-0.3cm}
\section{Conclusion}
In this letter, we proposed an MSP approach that leverages movable antennas and superdirective pairs to enhance mmWave communication performance without the need for phase shifters and attenuators. By optimizing the rotation angles and positions of antenna pairs, the MSP approach maximizes the received SNR in a multipath environment while reducing system complexity and hardware costs. An efficient algorithm based on alternating optimization and the Adam optimizer was developed to solve the non-convex optimization problem. 
Simulation results demonstrated that the proposed MSP approach achieves higher spectral efficiency compared to traditional FPA employing MRC. The MSP approach also avoids the insertion losses associated with RF components, further improving practical performance. 

\vspace{-0.3cm}

\bibliographystyle{IEEEtran}

    \bibliography{bibtex/bib/IEEEexample}

\end{document}